# PhyCV: The First Physics-inspired Computer Vision Library

Yiming Zhou, Callen MacPhee, Madhuri Suthar, and Bahram Jalali
Electrical and Computer Engineering Department
University of California, Los Angeles

*Abstract* - PhyCV is the first computer vision library which utilizes algorithms directly derived from the equations of physics governing physical phenomena. The algorithms appearing in the current release emulate, in a metaphoric sense, the propagation of light through a physical medium with natural and engineered diffractive properties followed by coherent detection. Unlike traditional algorithms that are a sequence of hand-crafted empirical rules or deep learning algorithms that are usually data-driven and computationally heavy, physics-inspired algorithms leverage physical laws of nature as blueprints for inventing algorithms. PhyCV features low-dimensionality and high-efficiency, making it ideal for edge computing applications. We demonstrate real-time video processing on NVIDIA Jetson Nano using PhyCV. In addition, these algorithms have the potential to be implemented in real physical devices for fast and efficient computation in the form of analog computing. The open-sourced code is available at https://github.com/JalaliLabUCLA/phycv

## 1. Introduction

PhyCV is a new class of computer vision algorithms inspired by physics, the current release of PhyCV has three algorithms: Phase-Stretch Transform (PST) [1], Phase-Stretch Adaptive Gradient-Field Extractor (PAGE) [2, 3], and Vision Enhancement via Virtual diffraction and coherent Detection (VEViD) [4]. These algorithms are originated from the research on photonic time stretch, which is a hardware technique for ultrafast and single-shot data acquisition that exploits dispersion and, in its full-field form, coherent detection [5, 6, 7, 8]. The algorithms emulate the propagation of light through a 2D physical medium with natural and artificial diffractive properties followed by coherent (phase) detection. The diffractive medium will apply a phase kernel to the frequency domain of the image and convert a real-valued image into a complex function. After coherent detection, the output phase contains useful features of the input image. In other words, PhyCV leverages the knowledge of optical physics and adapts it to computational imaging [9]. It is important to note that in PhyCV algorithms, the phase induced by propagation is very small ($\leq 2\pi$).

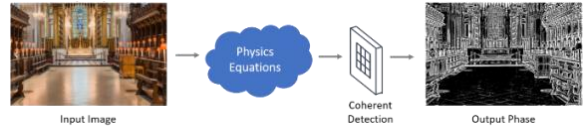

Figure 1: A conceptual diagram of PhyCV. The input image is transformed by physics equations in a physical system followed by coherent (phase) detection. The detected phase at the output contains useful features of the input image.

## 2. From Optical Physics to Algorithms

Photonic time stretch can be understood by considering the propagation of an optical pulse, on which the information of interest has been encoded, through a dispersive optical element. To outline how PhyCV emerged as an offshoot of photonic time stretch, we start with the Nonlinear Schrödinger Equation (NLSE) – the master equation that describes optical pulse propagation through an optical medium. The NLSE contains terms describing attenuation, dispersion and nonlinearity as shown below [10]:

$$\frac{\partial E(t,z)}{\partial z} = \frac{\alpha}{2}E(t,z) - i\frac{\beta_2}{2}\frac{\partial^2 E(t,z)}{\partial t^2} + i\gamma|E(t,z)|^2 E(t,z)$$

By disregarding the loss and nonlinearity (the first and the third terms), the NLSE can be simplified to the integral equation as shown below:

$$E_o(t) = \frac{1}{2\pi}\int \tilde{E}_i(\omega) \cdot e^{i\phi(\omega)} \cdot e^{i\omega t} d\omega$$

In photonic time-stretch systems, the information is encoded onto the spectrum of the input femtosecond pulse $\tilde{E}_i(\omega)$, through propagation, the spectrum is reshaped into a temporal signal at arbitrary time scale with a complex envelope $E_o(t)$ as shown in [6, 8].

Next, we convert the operation into discrete domain and introduce the 1D discrete stretch operator $\mathbb{S}$:

$$\mathbb{S}\{E_i[n]\} = IFFT\{FFT\{E_i(n)\} \cdot \tilde{K}(k_n) \cdot \tilde{L}(k_n)\}$$

Here, instead of using continuous variable $t$ and $\omega$, we use discrete variable $n$ and the corresponding frequency variable $k_n$. The 1D discrete stretch operator $\mathbb{S}$ can be written as the input signal spectrum multiplied by some general phase kernel $\tilde{K}(k_n)$ which is caused by dispersion or diffraction, and amplitude kernel $\tilde{L}(k_n)$. Next, we extend the stretch operator $\mathbb{S}$ to 2D discrete domain:

$$\mathbb{S}\{E_i[m,n]\} = IFFT^2\{FFT^2\{E_i(m,n)\} \cdot \tilde{K}(k_m, k_n) \cdot \tilde{L}(k_m, k_n)\}$$

Here $m$ and $n$ are spatial coordinates in the digital image, $k_m$ and $k_n$ are the corresponding frequency coordinates. This 2D discrete stretch operator emulates an image propagating through a metaphoric diffractive medium just like a laser pulse propagating through a dispersive optical element, which is the central process that governs photonic time stretch. This demonstrates how the algorithms are inspired by optical physics. The output of the 2D discrete stretch operator $\mathbb{S}$ is a complex-valued function. After doing coherent detection by mixing the output signal with a local oscillator (LO), the phase which contains useful feature of the input image is detected.

## 3. Algorithms in PhyCV

In this section, we introduce the three algorithms in the current release of PhyCV: Phase-Stretch Transform (PST), Phase-Stretch Adaptive Gradient-Field Extractor (PAGE), and Vision Enhancement via Virtual diffraction and coherent Detection (VEViD).

### 3.1. Phase-Stretch Transform (PST)

PST is a computationally efficient edge and texture detection algorithm with exceptional performance in visually impaired images [1, 11]. The mathematical operation of PST is shown below:

$$\mathbb{S}\{E_i[m,n]\} = IFFT^2\{FFT^2\{E_i(m,n)\} \cdot \tilde{K}(k_m, k_n) \cdot \tilde{L}(k_m, k_n)\}$$

$$PST\{E_i[m,n]\} = \sphericalangle\{\mathbb{S}\{E_i[m,n]\}\}$$

Here $E_i(m,n)$ is the input image, $m$ and $n$ are spatial coordinates, $k_m$ and $k_n$ are corresponding frequency coordinates. $\tilde{L}(k_m, k_n)$ is a Gaussian low-pass filter in the frequency domain for image denoising, $\tilde{K}(k_m, k_n) = e^{-i\phi(k_m, k_n)}$ is a nonlinear frequency-dependent phase filter which applies higher amount of phase to higher frequency features of the image. Since sharp transitions, such as edges and corners, contain higher frequencies, by detecting the phase of the output, PST extracts the edge information. The extracted edges are further enhanced by thresholding and morphological operations. In the implementation, we use a phase kernel $\tilde{K}(k_m, k_n)$ with a phase profile $\phi(k_m, k_n)$ that is symmetric in the polar coordinates as described below:

$$\phi(k_m, k_n) = \phi_{polar}(r, \theta) = \phi(r)$$

To create a low dimensional phase kernel with the required properties to perform edge and texture detection of the image, we choose the phase profile which has the derivative equals to the inverse tangent function:

$$\frac{d\phi(r)}{dr} = \tan^{-1}(r)$$

$$\phi(r) = r \cdot \tan^{-1}(r) - \frac{1}{2}\log(r^2 + 1)$$

Therefore, the PST kernel is implemented as:

$$\phi(k_m, k_n) = \phi(r)$$
$$= S \cdot \frac{Wr \cdot \tan^{-1}(Wr) - \frac{1}{2}\log(1 + (Wr)^2)}{Wr_{max} \cdot \tan^{-1}(Wr_{max}) - \frac{1}{2}\log(1 + (Wr_{max})^2)}$$

PST can also be implemented with other phase kernels. The necessary properties of the kernel have been described in [11].

PST has been applied to various tasks including improving the resolution of MRI image [12], extracting blood vessels in retina images to identify various diseases [13], detection of dolphins in the ocean [14], waste water treatment [15], single molecule biological imaging [16], and classification of UAV using micro imaging [17].

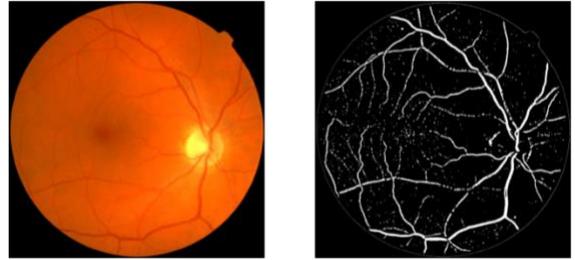

Figure 2: Retina vessel detection using PST in PhyCV.

### 3.2. Phase-Stretch Adaptive Gradient-Field Extractor (PAGE)

PAGE is a physics inspired feature engineering algorithm that computes a feature set comprised of edges at different spatial frequencies (and hence spatial scales) and orientations [2, 3]. Metaphorically speaking, PAGE emulates the physics of birefringent (orientation-dependent) diffractive propagation through a physical medium with a specific diffractive

structure. The mathematical operation of PAGE is shown below:

$$\mathbb{S}\{E_i[m,n];\theta\} = IFFT^2\{FFT^2\{E_i(m,n)\} \cdot \widetilde{K}(k_m,k_n;\theta) \cdot \widetilde{L}(k_m,k_n)\}$$

$$PAGE\{E_i[m,n];\theta\} = \measuredangle\{\mathbb{S}\{E_i[m,n];\theta\}\}$$

Here $E_i(m,n)$ is the input image and $\widetilde{L}(k_m,k_n)$ is the denoising filter. Then instead of having one phase filter, PAGE has the phase filter bank $\widetilde{K}(k_m,k_n;\theta)$, which contains filters with different angle variable $\theta$ that controls the directionality of the detected edge. A change of basis leads to the transformed frequency variables $k_m'$ and $k_n'$:

$$k_m' = k_m \cdot \cos\theta + k_n \cdot \sin\theta$$

$$k_n' = -k_m \cdot \sin\theta + k_n \cdot \cos\theta$$

The PAGE kernel $\widetilde{K}(k_m,k_n;\theta)$ now becomes $\widetilde{K}(k_m',k_n')$ and it is expressed as a product of two phase functions, $\phi_1$ and $\phi_2$. The first component $\phi_1$ is a symmetric gaussian filter that selects the spatial frequency range of the edges that are detected. Default center frequency is 0, which indicates a baseband filter, the center frequency and bandwidth of which can be changed to probe edges with different sharpness. In other words, it enables the filtering of edges occurring over different spatial scales. The second component, $\phi_2$, performs the edge-detection. The explanation of the parameters can be found in [2].

$$\widetilde{K}(k_m,k_n;\theta) = \widetilde{K}(k_m',k_n') = \exp(-i \cdot \phi_1(k_m') \cdot \phi_2(k_n'))$$

$$\phi_1(k_m') = S_1 \cdot \frac{1}{\sqrt{2\pi}\sigma_1} \cdot \exp\left(-\frac{(|k_m'|-\mu_1)^2}{2\sigma_1^2}\right)$$

$$\phi_2(k_n') = S_2 \cdot \frac{1}{\sqrt{2\pi}|k_n'|\sigma_2} \cdot \exp\left(-\frac{(ln|k_n'|-\mu_2)^2}{2\sigma_2^2}\right)$$

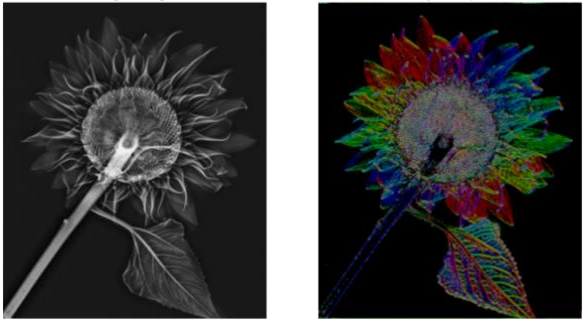

Figure 3: Directional edge detection of a sunflower using PAGE in PhyCV. For visualization, the directions of the edges are mapped into colors.

## 3.3. Vision Enhancement via Virtual diffraction and coherent Detection (VEViD)

VEViD is an efficient and interpretable low-light and color enhancement algorithm that reimagines a digital image as a spatially varying metaphoric light field and then subjects the field to the physical processes akin to diffraction and coherent detection [4]. The term "virtual" captures the deviation from the physical world. The light field is "pixelated" and the propagation imparts a phase with an arbitrary dependence on frequency which can be different from the quadratic behavior of physical diffraction. The mathematical operation of VEViD is shown below:

$$\mathbb{S}\{E_i[m,n;c]\} = IFFT^2\{FFT^2\{E_i(m,n;c) + b\} \cdot \widetilde{K}(k_m,k_n)\}$$

$$VEViD\{E_i[m,n;c]\} = \tan^{-1}\left(G \cdot \frac{Im\{\mathbb{S}\{E_i[m,n;c]\}\}}{E_i[m,n;c]}\right)$$

Here $E_i[m,n;c]$ is the input digital image and $c$ represents the color channel in the HSV color space. VEViD leads to low-light enhancement when operating on V (value) channel as shown in Figure 3 and color enhancement when operating on S (saturation) channel as shown in Figure 4. $b$ is a regularization term and $G$ is the phase activation gain term. $\widetilde{K}(k_m,k_n)$ is the phase kernel which has a phase profile $\phi(k_m,k_n)$ that follows a Gaussian distribution with zero mean.

$$\widetilde{K}(k_m,k_n) = \exp(-i \cdot \phi(k_m,k_n))$$

$$\phi(k_m,k_n) = S \cdot \exp(-\frac{k_m^2 + k_n^2}{T})$$

VEViD can be further accelerated through mathematical approximations that reduce the computation time without appreciable sacrifice in image quality. A closed-form approximation for VEViD which we call VEViD-lite is shown below and can achieve up to 200 FPS for 4K video enhancement. The full derivation of the physical and mathematical principle of VEViD can be found in [4].

$$VEViD_{lite}\{E_i[m,n;c]\} = \tan^{-1}\left(-G \cdot \frac{E_i[m,n;c]+b}{E_i[m,n;c]}\right)$$

It is also demonstrated that the VEViD serves as a powerful pre-processing tool that improves neural network based object detector in night-time environments without adding computational overhead and retraining.

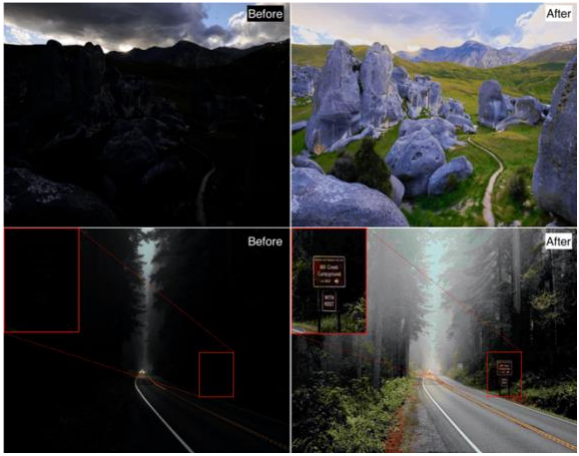

Figure 3: Low-light enhancement using VEViD in PhyCV.

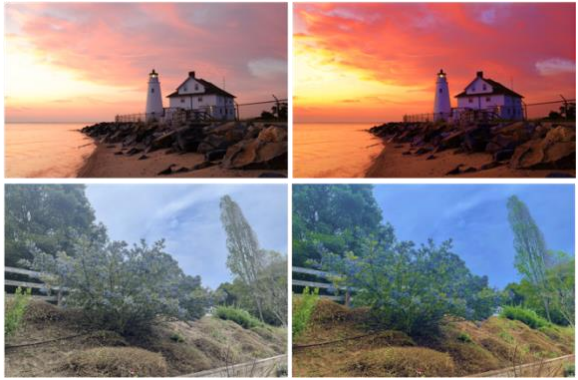

Figure 4: Color enhancement using VEViD in PhyCV.

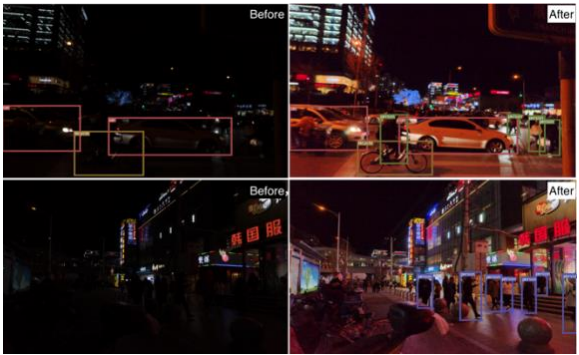

Figure 5: The performance of YOLO-v3 object detector is improved by VEViD in PhyCV.

## 4. PhyCV on the Edge

Featuring low-dimensionality and high-efficiency, PhyCV is ideal for edge computing applications. In this section, we demonstrate running PhyCV on NVIDIA Jetson Nano in real-time.

### 4.1. NVIDIA Jetson Nano Developer Kit

NVIDIA Jetson Nano Developer Kit is a small-sized and power-efficient platform for edge computing applications. It is equipped with an NVIDIA Maxwell architecture GPU with 128 CUDA cores, a quad-core ARM Cortex-A57 CPU, 4GB 64-bit LPDDR4 RAM, and supports video encoding and decoding up to 4K resolution. Jetson Nano also offers a variety of interfaces for connectivity and expansion, making it ideal for a wide range of AI and IoT applications. In our setup, we connect a USB camera to the Jetson Nano to acquire videos and demonstrate using PhyCV to process the videos in real-time.

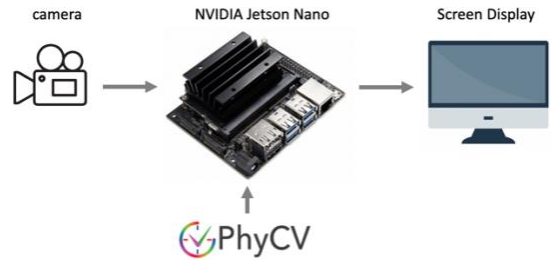

Figure 6: Conceptual diagram of running PhyCV on Jetson Nano for real-time video processing.

### 4.2. Real-time PhyCV on Jetson Nano

We use the Jetson Nano (4GB) with NVIDIA JetPack SDK version 4.6.1, which comes with pre-installed Python 3.6, CUDA 10.2, and OpenCV 4.1.1. We further install PyTorch 1.10 to enable the GPU accelerated PhyCV. We demonstrate the results and metrics of running PhyCV on Jetson Nano in real-time for edge detection and low-light enhancement tasks in Figure 7 and Table 1. For 480p videos, both operations achieve beyond 38 FPS, which is sufficient for most cameras that capture videos at 30 FPS. For 720p videos, PhyCV low-light enhancement can operate at 24 FPS and PhyCV edge detection can operate at 17 FPS.

We compare runtimes for low-light enhancement between PhyCV and one of the leading neural networks called Zero-DCE [18] as shown in Figure 8. PhyCV operates at 24 FPS while Zero-DCE is even slower than 1 FPS. It's important to note that both implementations are done in Python with OpenCV and PyTorch without any other optimization. We expect to see further acceleration when implementing PhyCV with C/C++ and integrating it with dedicated NVIDIA SDKs for Jetsons.

|  | 480p | 720p |
|---|---|---|
| Edge Detection | 25.9 ms | 58.5 ms |
| Low-light Enhancement | 24.5 ms | 41.1 ms |

Table 1: Running time (per frame) of PhyCV on Jetson Nano for videos at different resolutions.

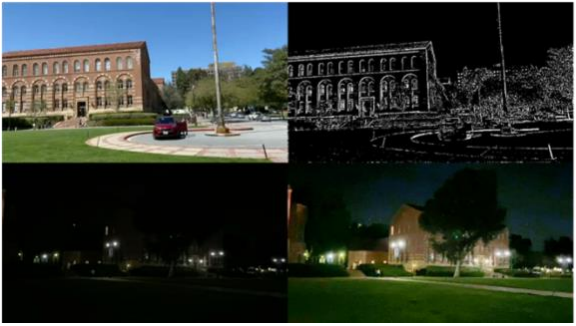

Figure 7: Running PhyCV in real-time on Jetson Nano. Top: real-time edge detection. Bottom: real-time low-light enhancement. You can find the corresponding videos at: https://en.wikipedia.org/wiki/PhyCV.

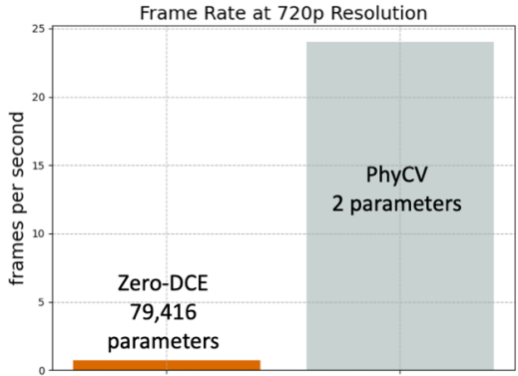

Figure 8: Frame rate comparison on the low-light enhancement task for a 720p video on Jetson Nano between Zero-DCE and PhyCV.

## 5. PhyCV Highlights

### 5.1. Modular Code Architecture

The modular code architecture of PhyCV follows the physics behind the algorithm and is therefore more intuitive. Since all algorithms in PhyCV emulate the propagation of the input image through a device with specific diffractive properties, which applies a phase kernel to the frequency domain of the original image. This process has three steps in general, loading the image, initializing the kernel and applying the kernel. In the implementation, each algorithm is represented as a class and each class has methods that simulate the steps described above. This makes the code easy to understand and extend.

```
class PST:
    def __init__(self, ...):

    def load_img(self, ...):

    def init_kernel(self, ...):

    def apply_kernel(self, ...):

    def run(self, ...):
```

Figure 9: The modular code architecture of PhyCV algorithm PST.

### 5.2. GPU Acceleration

PhyCV also supports GPU acceleration. The GPU versions of PhyCV are built on PyTorch and accelerated by CUDA. The GPU compatibility significantly accelerates the algorithms, which is beneficial for real-time video processing and related deep learning tasks. Here we show the comparison of running time of PhyCV algorithms on CPU and GPU for videos at 1080p, 2K and 4K resolutions. For results shown in the table below, the CPU is Intel i9-9900K @ 3.60GHz x 16 and the GPU is NVIDIA TITAN RTX. Note that the PhyCV low-light enhancement operates in the HSV color space, so the running time also includes RGB to HSV conversion. Moreover, for running time using GPUs, we ignore the time of moving data from CPUs to GPUs and count the algorithm operation time only.

|  | CPU | GPU |
|---|---|---|
| 1080p | 550 ms | 4.6 ms |
| 2K | 1000 ms | 8.2 ms |
| 4K | 2290 ms | 18.5 ms |

Table 2. Running time (per frame) of PhyCV – PST edge detection on videos at different resolutions.

|  | CPU | GPU |
|---|---|---|
| 1080p | 2800 ms | 48.5 ms |
| 2K | 5000 ms | 87 ms |
| 4K | 11660 ms | 197 ms |

Table 3. Running time (per frame) of PhyCV – PAGE directional edge detection on videos at different resolutions.

|  | CPU | GPU |
|---|---|---|
| 1080p | 175 ms | 4.3 ms |
| 2K | 320 ms | 7.8 ms |
| 4K | 730 ms | 17.9 ms |

Table 4. Running time (per frame) of PhyCV – VEViD low-light enhancement on videos at different resolutions. RGB to HSV conversion time is included.

|        | CPU    | GPU    |
|--------|--------|--------|
| 1080p  | 60 ms  | 2.1 ms |
| 2K     | 110 ms | 3.5 ms |
| 4K     | 245 ms | 7.4 ms |

Table 5. Running time (per frame) of PhyCV – VEViD-lite low-light enhancement on videos at different resolutions. RGB to HSV conversion time is included.

**GitHub Repository**

https://github.com/JalaliLabUCLA/phycv